\documentclass[%
  reprint,
  superscriptaddress,
 amsmath,amssymb,
 aps,
]{revtex4-2}

\usepackage{graphicx}
\usepackage{dcolumn}
\usepackage{bm}
\usepackage{siunitx}

\begin{document}

\title{Experimental observation of the spatio-temporal dynamics of breather gases in a recirculating fiber loop}

\author{Fran\c{c}ois Copie}
\affiliation{Univ. Lille, CNRS, UMR 8523 - PhLAM - Physique des Lasers Atomes et Mol\'ecules, F-59 000 Lille, France}
\author{Gino Biondini}
\affiliation{Department of Mathematics, State University of New York, Buffalo, NY 14260}
\author{Jeffrey Oregero}
\affiliation{Department of Mathematics, University of Kansas, Lawrence, KS 66045}
\author{Gennady El}
\affiliation{Department of Mathematics, Physics and Electrical Engineering, Northumbria University, Newcastle upon Tyne, NE1 8ST, United Kingdom}
\author{Pierre Suret}
\affiliation{Univ. Lille, CNRS, UMR 8523 - PhLAM - Physique des Lasers Atomes et Mol\'ecules, F-59 000 Lille, France}
\author{St\'ephane Randoux}
\affiliation{Univ. Lille, CNRS, UMR 8523 - PhLAM - Physique des Lasers Atomes et Mol\'ecules, F-59 000 Lille, France}

\date{\today}

\begin{abstract}
  We report the first experimental observation of breather gases (BGs) in optics, realized in a recirculating fiber loop enabling virtually lossless propagation over $1200$ km. Initiated by a slowly modulated optical background perturbed by noise, the BGs form through a nonlinear fission process and demonstrate spatiotemporal dynamics that align closely with numerical simulations of the focusing one-dimensional nonlinear Schrödinger equation.  The minimal dissipation in our setup enables a statistical characterization of the BGs and confirms the theoretically predicted doubling of kurtosis during the evolution of the BGs. These results open new avenues for experimental studies of integrable turbulence involving solitons on finite background.
\end{abstract}

\maketitle


Solitons are stable, localized wavepackets that propagate without changing shape, arising from a precise balance between nonlinear and dispersive effects in a material medium. These waves are ubiquitous in nature and play a fundamental role in various scientific and technological fields, particularly in nonlinear fiber optics \cite{agrawal2019nonlinear}. Mathematically, solitons correspond to specific solutions of certain partial differential equations, such as the one-dimensional nonlinear Schrödinger equation (1D-NLSE), which  is a paradigmatic model for the description of nonlinear dispersive waves dynamics in various physical systems, including water waves, matter waves and optical fibers. 

Beyond conventional solitons, which are localized structures on a zero background, the focusing 1D-NLSE admits a broader class of solutions known as breathers, or solitons on finite background. This category includes the celebrated Akhmediev, Peregrine, and Kuznetsov-Ma breathers, whose complex dynamics have been experimentally investigated in fiber optics,  hydrodynamics and Bose-Einstein condensates \cite{Kibler:10, Hammani:11a, Dudley:14, Frisquet:13, Kibler:15, Gelash:22, Chabchoub:12a, Xu:19, Xu:19b, Romero:24}.

While the classical theory of the focusing 1D-NLSE primarily deals with regular, deterministic soliton and breather structures \cite{Zakharov:13,Biondini:14,GEl:16}, recent studies in integrable turbulence \cite{Soto:16,Congy:24} have emphasized the importance of {\it random} nonlinear wave fields, such as soliton gases (SGs). A SG consists of a large ensemble of interacting solitons with randomly distributed amplitudes, velocities, and positions \cite{Zakharov:71,GEl:05,GEl:20,Suret:20,Suret:23,Suret:24}. Notably, SG dynamics has been shown to underlie fundamental physical phenomena such as spontaneous modulational instability (MI) \cite{Gelash:19} and the emergence of rogue waves in focusing media \cite{Gelash:18}. 

As recently introduced in \cite{GEl:20}, breather gases (BGs) generalize the concept of SGs to random ensembles of solitons on finite-amplitude backgrounds. Although the numerical synthesis of BGs has been reported in Ref.\,\cite{Roberti:21}, BGs have not yet been observed in experiments. Recent theoretical studies have indicated however that BGs can emerge spontaneously from a fission process arising in the nonlinear evolution of slowly modulated, periodic backgrounds perturbed by small initial noise \cite{Biondini:25}.

Building on this concept, we present optical fiber experiments in which we observe and investigate the spatiotemporal dynamics of BGs. The primary challenge in these experiments is dissipation, as BGs require it to be minimal for the conservation of the amplitude of the background. We have successfully achieved this condition in a recirculating fiber loop, enabling virtually lossless nonlinear propagation over approximately \SI{1200}{km}. This represents a significant improvement compared to previous works on MI, higher-order MI, Fermi-Pasta-Ulam-Tsingou recurrences or breathers, where dissipation, though small, notably influenced the dynamics \cite{Erkintalo:11, Hammani:11b, Kraych:19a, Kraych:19b, Coppini:20, Copie:22, Mussot:14, Mussot:18, Vanderhaegen:24}. This advancement allows us to explore the spatiotemporal dynamics of BGs in single-shot and in particular, to check theoretical predictions about kurtosis doubling in the evolution of BGs \cite{Biondini:25}.

\vspace{0.4cm}

\begin{figure}[ht]
\centering
\includegraphics[width=\linewidth]{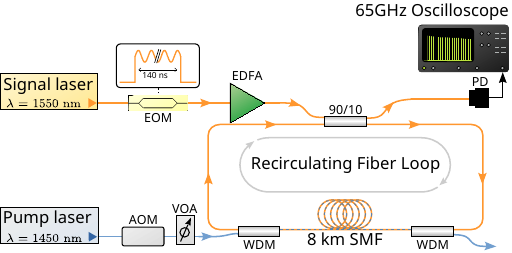}
\caption{Schematic of the experimental setup. Initially, the BG takes the shape of a \SI{140}{ns}-long square pulse at \SI{1550}{nm}, with its power slowly modulated at \SI{250}{MHz}. The modulated pulse, perturbed by optical noise from an Erbium-doped fiber amplifier (EDFA), propagates through a recirculating fiber loop. The optical gain rate within the loop is precisely controlled via Raman amplification using a \SI{1450}{nm} pump laser to compensate the losses accumulated over a roundtrip.
}
\label{fig1}
\end{figure}

Figure 1 illustrates our experimental setup, which features a recirculating fiber loop constructed from approximately \SI{8}{km} of single-mode fiber (SMF). The loop is closed using a $90/10$ fiber coupler, recirculating $90\%$ of the optical power. The optical signal travels clockwise, and at each roundtrip, $10 \%$ of the circulating power is extracted and directed to a photodetector (PD) connected to a fast oscilloscope with a sampling rate of \SI{160}{GSa/s} and a bandwidth of \SI{65}{GHz}. The combined detection bandwidth of the oscilloscope and PD is \SI{32}{GHz}. The periodic extraction of light from the loop allows for stroboscopic monitoring of the wavefield's evolution every \SI{8}{km}. The recorded signals are subsequently processed numerically to generate space-time diagrams illustrating the wavefield dynamics over hundreds of roundtrips within the fiber loop \cite{Kraych:19a, Kraych:19b, Suret:23}.

In our experiment, the initial condition consists of a \SI{140}{ns}-long square pulse at \SI{1550}{nm} with a power profile that is not constant over time. Instead, it is slowly modulated sinusoidally with a period of $T_m= \SI{4}{ns}$, which is significantly longer than the characteristic duration of the breather structures that emerge during the nonlinear evolution (typically \SI{50}{ps} for the range of parameter of the experiments). This modulated square pulse is generated using a fast electro-optic modulator (EOM) and is intentionally perturbed by a small amount of optical noise, introduced by an Erbium-doped fiber amplifier (EDFA), before injection into the fiber loop (see Fig.\,1). Following the scenario proposed in Ref.\,\cite{Biondini:25}, the modulated square pulse, perturbed by this optical noise, evolves into a fully randomized BG within the fiber loop, see Fig.\,2(a). The perturbation of the initially modulated field by noise is a crucial requirement for the randomization of the BG that would otherwise exhibit a regular dynamics \cite{Erkintalo:11, Sheveleva:22, Vanderhaegen:22} which can be interpreted to some extent in terms of local breather states \cite{Erkintalo:11b}. 

\begin{figure}[ht]
\centering
\includegraphics[width=\linewidth]{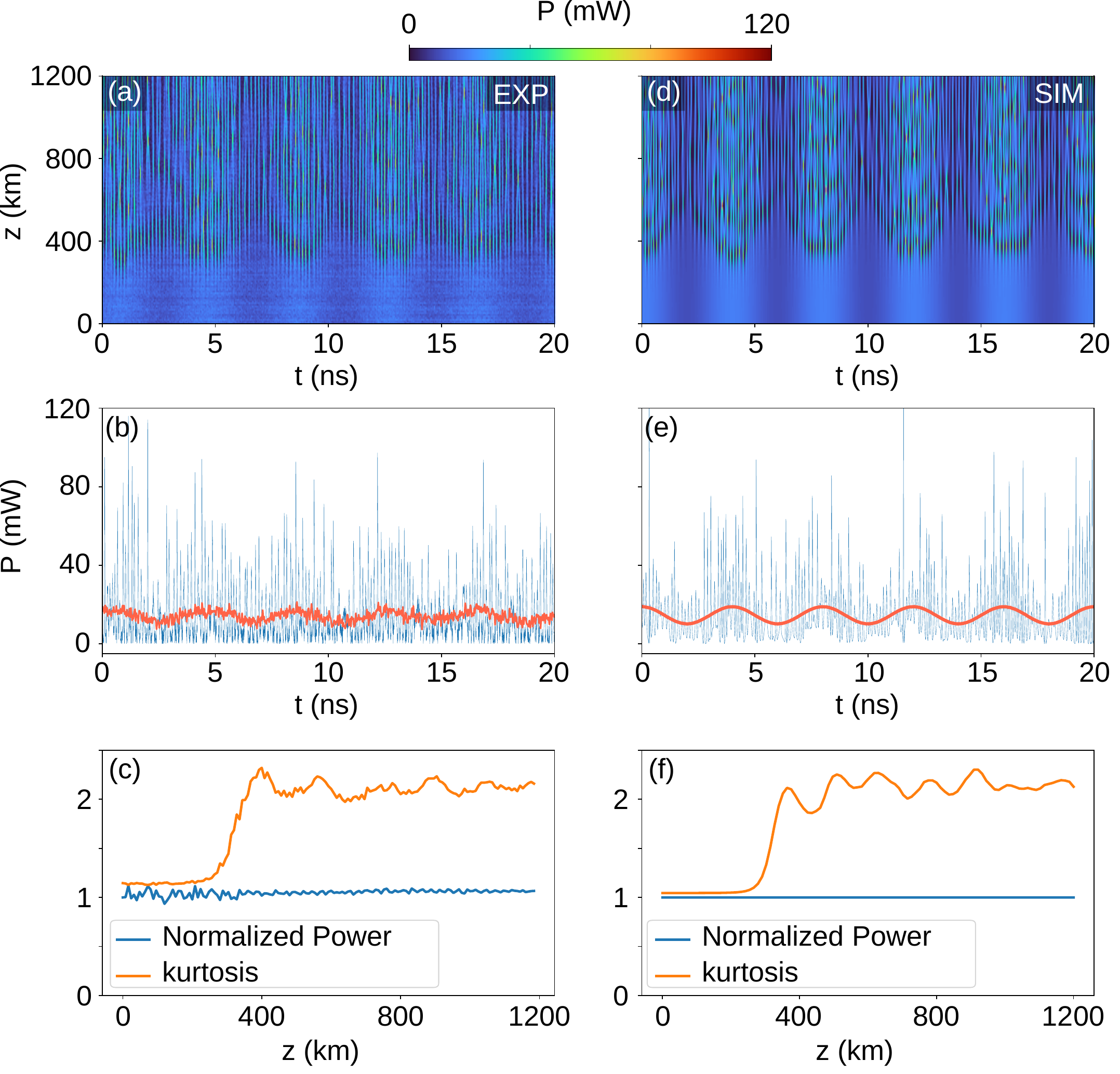}
\caption{ (a) Space-time evolution of the BG recorded in the recirculating fiber loop over a propagation distance of \SI{1200}{km}. The BG emerges from a fission process where the noisy optical background with a slowly-modulated power at $f_m= \SI{250}{MHz}$ is destabilized after a propagation distance of $\sim \SI{300}{km}$. (b) Power of the optical field recorded at $z=\SI{0}{km}$ (red line) and $z=\SI{1200}{km}$ (blue line). (c) Experimental evolution of the normalized optical power (blue line) and of the kurtosis (orange line) in the recirculating fiber loop. (d), (e), (f) Same as (a), (b), (c) but from numerical simulations of Eq.\,(\ref{eq:NLSE}) with the initial condition given by Eq.\,(\ref{eq:CI}) for $\Delta T= \SI{140}{ns}$, $f_m=\SI{250}{MHz}$, $P_0=\SI{14.5}{mW}$, $m=0.3$.
}
\label{fig2}
\end{figure}

As shown in Fig. 1, the SMF with a coefficient $\beta_2 = \SI{-22}{ps^2/km}$ for group velocity dispersion and a Kerr nonlinearity coefficient $\gamma = \SI{1.3}{\per\watt\per\kilo\meter}$ is connected at both ends to wavelength division multiplexers (WDMs). These WDMs are used to inject and extract the light of a \SI{1450}{nm} pump laser into and out of the fiber loop, thereby achieving backward Raman amplification of the optical signal at \SI{1550}{nm}. Importantly, the power of the \SI{1450}{nm} laser is gated using an acousto-optic modulator (AOM) to match the required number of roundtrips ($150$ in Fig.\,2 and 3) within the fiber loop. Additionally, the optical power at \SI{1450}{nm} is finely adjusted using a variable optical attenuator (VOA) to compensate for the total dissipation experienced by the \SI{1550}{nm} signal during a roundtrip. In this way, effective losses are nearly canceled for the \SI{1550}{nm} signal, allowing it to propagate over \SI{1200}{km} ($150$ roundtrips) with a nearly constant mean power, as illustrated in Fig.\,2(c). In our experiments, the mean power of the modulated square pulse propagating in the recirculating fiber loop is typically around \SI{15}{mW}.  

Figures 2(a, b) demonstrate that the dynamical features predicted theoretically in Ref.\,\cite{Biondini:25} are qualitatively observed in our experiment. The slowly-modulated background, initially perturbed by a small amount of optical noise, undergoes a destabilization (or fission \cite{Biondini:25}) that becomes clearly observable after a propagation distance of approximately \SI{300}{km}, as shown in Fig.\,2(a). At long propagation distances, the optical field consists of a random ensemble of coherent structures with a duration of approximately \SI{50}{ps}, which is much shorter than the period (\SI{4}{ns}) of the initial modulated field, see Fig.\,2(a).

As shown in Fig.\,2(d, e), the dynamical features observed in our experiment are well reproduced by numerical simulations of the 1D-NLSE:
\begin{equation}\label{eq:NLSE}
  i\frac{\partial A}{\partial z}=\frac{\beta_2}{2}\frac{\partial^2 A}{\partial t^2}-\gamma|A|^2A ,
\end{equation}
where $A(z,t)$ represents the complex envelope of the electric field that slowly varies in physical space $z$ and time $t$. Figure 2(d, e) show the evolution of the optical power $P(z,t)$ defined as the modulus square of the envelope of the optical field: $P(z,t)= |A(z,t)|^2$. The initial condition in our numerical simulation replicates the profile of the modulated square pulse used in the experiments which reads:
\begin{equation}\label{eq:CI}
  A(z=0,t) = \sqrt{P_0 \, (1 + m \, \cos (2\pi f_m t ) )} \quad e^{-(2t/\Delta T)^{2p}}   + \zeta(t) ,
\end{equation}
where $P_0=\SI{14.5}{mW}$ represents the mean power of the modulated square pulse that has a full width at half maximum $\Delta T=\SI{140}{ns}$. The modulation frequency $f_m=1/T_m$ is \SI{250}{MHz} like in the experiment, $m$ represents a dimensionless modulation index which is equal to $0.3$ in the numerical simulations reported in Fig.\,2, and $p$ is an integer that determines the rise and fall times of the square pulse. In our numerical simulations, its value is set to $p=10$. 
The term $\zeta(t)$ represents a small noise term whose strength and correlation time are determined from the Fourier power spectrum of the initial condition measured in the experiment. The exact properties of the noise $\zeta(t)$ do not notably influence the scenario of the BG formation. Our numerical simulations are performed using an $8$th-order adaptive step Runge-Kutta method in a numerical box of size \SI{300}{ns}, discretized with $65 536$ points.

Figures 3(a, b) and (d, e) show qualitatively the same experimental and numerical features as Fig.\,2(a, b) and (d, e), but with a larger modulation index value of $m=0.6$ and a slightly larger mean power $P_0=\SI{15}{mW}$. Due to these larger values, the formation of the BG begins at a slightly shorter propagation distance of approximately \SI{300}{km}, compare Fig.\,2(a) with Fig.\,3(a). 

\begin{figure}[ht]
\centering
\includegraphics[width=\linewidth]{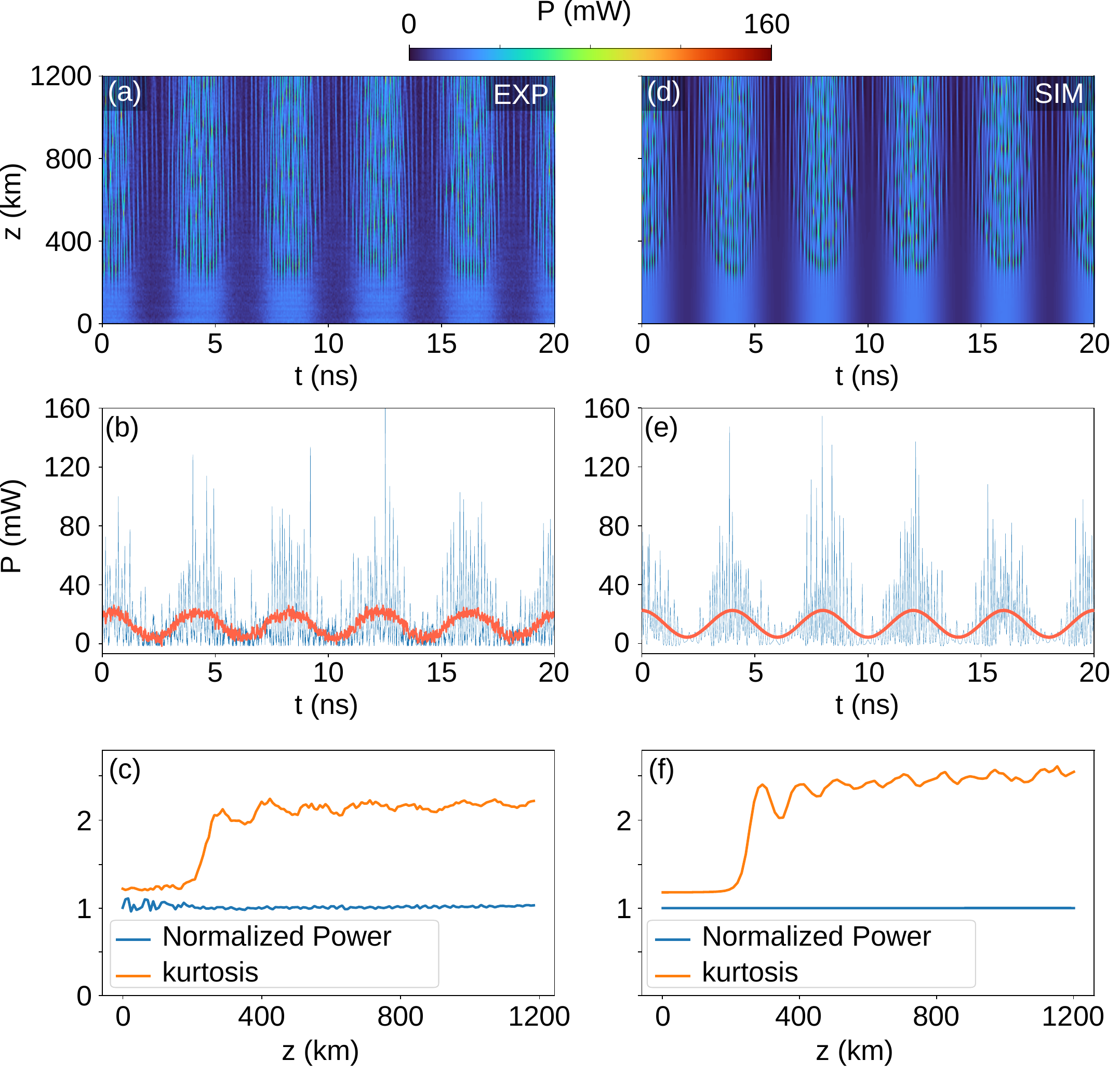}
\caption{Same as Fig.\,2 but with $P_0=\SI{15}{mW}$, $m=0.6$.
}
\label{fig3}
\end{figure}

Given the nearly constant optical power throughout the propagation distance in our experiment, as shown in Fig.\,2(c) and 3(c), we are able to perform a statistical analysis of the BG's properties following its formation from the fission of the initial modulated wave. This analysis allows us to verify the theoretical prediction of Ref.\,\cite{Biondini:25} that the fourth-order moment of the BG doubles during its evolution. The normalized fourth-order moment of the BG, often referred to as the kurtosis, is defined by:
\begin{equation}\label{eq:kurtosis}
\kappa(z) = T_{BG} \frac{ \int_0^{T_{BG}} P^2(z,t) dt} { \left( \int_0^{T_{BG}} P(z,t) dt \right)^2} 
\end{equation}
where $T_{BG}$ represents the width of the time window over which the statistical analysis of the BG is conducted. 
To prevent edge effects from affecting the statistical analysis -- such effects are associated to the formation of dispersive shock waves at the sharp edges of the square pulse \cite{GEl:16, Bonnefoy:20} -- we conduct the analysis at the center of the square pulse. We set $T_{BG}=\SI{120}{ns}$, a duration that is shorter than the pulse length $\Delta T=\SI{140}{ns}$ but significantly longer than the modulation period $T_m=\SI{4}{ns}$.

Notably, the theoretical analysis in \cite{Congy:24, TovbisWang:22} predicts 
a doubling of the initial value of kurtosis in the process of the long-distance propagation, resulting in $\kappa_\infty = 2 \kappa(0)$, which is a general feature of the SG/BG fission of partially coherent waves. As suggested by the results of Ref.~\cite{Biondini:25}, this doubling process also applies to the current context of the evolution  of the noise-augmented real-valued periodic input data (\eqref{eq:CI}) with large period. Similar to Ref.~\cite{Biondini:25}, the role of the initial kurtosis $\kappa(0)$ is now played by the fourth normalized moment of the  (deterministic) input profile given by Eq.~\ref{eq:CI} with $\zeta(t)=0$.
In particular, if one neglects the exponential windowing function, the profile in Eq.~\ref{eq:CI} gives rise to simple trigonometric integrals, which can be readily computed analytically to obtain $\kappa(0) = 1 + m^2/2$.

In the first experiment, with the lowest modulation index $m=0.3$, Fig.\,2(c) shows that the kurtosis value $\kappa(z)$ remains nearly constant at approximately $1.07$ during the initial \SI{300}{km} of propagation. The fission process, which leads to the formation of the optical BG, occurs between approximately \SI{300}{km} and \SI{400}{km}, where the kurtosis value abruptly doubles. By $z=\SI{1200}{km}$, the final value of $\kappa(z)$reaches approximately $2.0$, nearly twice its initial value. Qualitatively similar features are observed for the second optical BG, which was realized using a larger modulation index $m$ of $0.6$, as shown in Fig.\,3(c). The initial value of the kurtosis is slightly higher ($\kappa(0<z<\SI{100}{km}) \simeq 1.21$) than in Fig.\,2(c), due to the larger initial modulation index. Figure 3(c) indicates a fission phenomenon occurring at a shorter propagation distance around $\sim \SI{200}{km}$, as previously noted in the description of the space-time evolution of this BG. At long propagation distance ($z \sim \SI{1200}{km}$), the kurtosis value $\kappa$ has approximately doubled, reaching $\sim 2.18$. 

Figures 2(f) and 3(f) illustrate that the evolution of the kurtosis along the propagation distance $z$ is nearly identical in numerical simulations of Eq.\,(\ref{eq:NLSE}), using the initial condition specified by Eq.\,(\ref{eq:CI}). We note that the kurtosis in numerical simulations reaches values that are slightly larger than the experimental ones, compare Fig.\,2(c) with Fig.\,2(f) and Fig.\,3(c) with Fig.\,3(f). As discussed in detail in Ref.\,\cite{Kraych:19b}, this phenomenon is due to the finite detection bandwidth of \SI{32}{GHz} in the experiments, which is not accounted for in the numerical simulations of the 1D-NLSE presented in Fig.\,2(f) and 3(f).

\begin{figure}[ht]
\centering
\includegraphics[width=\linewidth]{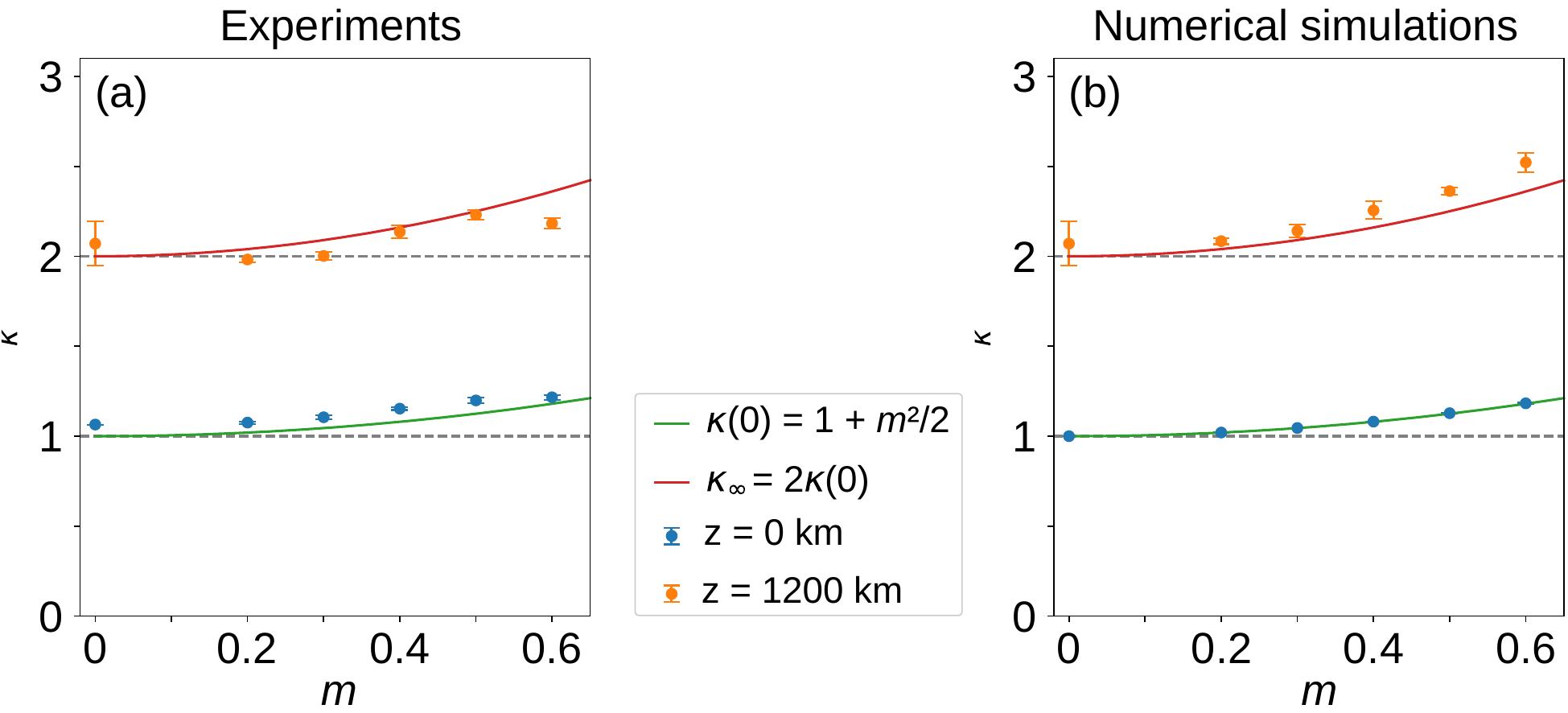}
\caption{ (a) Experiments. Evolution of the kurtosis $\kappa$ defined by Eq. (\ref{eq:kurtosis}) as a function of the modulation index $m$ of the initial condition at $z=0$ km (blue points) and at $z=1200$ km (orange points). (b) Same as in (a) but from numerical simulations of Eq. (\ref{eq:NLSE}) using Eq. (\ref{eq:CI}) as initial condition.  The green (resp. orange) curve represents the function $\kappa(0) = 1 + m^2/2$ (resp. $\kappa_\infty = 2 \kappa(0)$).
}
\label{fig4}
\end{figure}

To further validate the statistical analysis of the BGs, Figure 4(a) displays the experimentally measured kurtosis values $\kappa$ at $z\sim{0}$ km (blue points) and at $z=\sim{1200}$ km (orange points) for six different values of the modulation index $m$, ranging from $0$ to $0.6$. The results are presented as mean values calculated over 120 km intervals, with error bars representing the standard deviation of the kurtosis fluctuations in these intervals. As it can be anticipated from Ref.\,\cite{Biondini:25}, the initial kurtosis $\kappa(0)$ increases monotonically according to $\kappa(0) = 1 + m^2/2$, as indicated by the green line in Fig.~4(a). At $z=\SI{1200}{km}$, the statistical analysis of our experimental data shows that the kurtosis value has approximately doubled for all the values of $m$. This behavior is consistent with numerical simulations of Eq.\,(\ref{eq:NLSE}), as shown in Fig.\,4(b). Note that numerical simulations, which are unaffected by limitations arising from the finite detection bandwidth, depict kurtosis values that are slightly higher than those obtained experimentally. Furthermore, numerical simulations show that at a propagation distance of $1200$ km, the kurtosis has not yet stabilized and is slightly higher than twice its initial value. These simulations suggest that, with our experimental parameters, the kurtosis would exactly double after a propagation distance of approximately 8000 km.

In conclusion, we have presented the first experimental observation of BGs in nonlinear fiber optics, enabled by a recirculating fiber loop that supports virtually lossless propagation over \SI{1200}{km}. Initiated from a slowly modulated optical background perturbed by noise, the BGs are formed through a nonlinear fission process and demonstrate spatiotemporal dynamics that align closely with numerical simulations of the focusing 1D-NLSE. Notably, the observed doubling of kurtosis during the evolution confirms key aspects of the theoretical framework proposed in Ref.\,\cite{Biondini:25}. These findings establish the physical relevance of BGs in optical fiber systems and open promising perspectives for experimental studies of integrable turbulence.

\section*{Acknowledgment}  This work has been partially supported  by the Agence Nationale de la Recherche  through the SOGOOD (ANR-21-CE30-0061) and StormWave (ANR-21-CE30-0009) projects, the LABEX CEMPI project (ANR-11-LABX-0007), the Ministry of Higher Education and Research, Hauts de France council and European Regional Development Fund (ERDF) through the Nord-Pas de Calais Regional Research Council and the European Regional Development Fund (ERDF) through the Contrat de Projets Etat-R\'egion (CPER Photonics for Society P4S). The authors would like to thank the Centre d'Etudes et de Recherche Lasers et Application (CERLA) for technical support and the Isaac Newton Institute for Mathematical Sciences, Cambridge, for support and hospitality during the programme ``Emergent phenomena in nonlinear dispersive waves'', where work on this paper was undertaken. This work was supported by EPSRC grant EP/V521929/1.



\bigskip



%

\end{document}